# Accelerated Lifetime Testing and Analysis of Delta-doped Silicon Test Structures

Connor Halsey[1], Jessica Depoy[1], DeAnna M. Campbell[1], Daniel R. Ward, Evan M. Anderson [1], Scott W. Schmucker[1], Jeffrey A. Ivie[1], Xujiao Gao[1], David A. Scrymgeour[1], and Shashank Misra[1]

*Abstract*— As transistor features shrink beyond the 2 nm node, studying and designing for atomic scale effects become essential. Being able to combine conventional CMOS with new atomic scale fabrication routes capable of creating 2D patterns of highly doped phosphorus layers with atomic precision has implications for the future of digital electronics. This work establishes the accelerated lifetime tests of such doped layers, showing that these materials survive high current (>3.0 MA/cm$^2$) and 300°C for greater than 70 days and are still electrically conductive. The doped layers compare well to failures in traditional metal layers like aluminum and copper where mean time to failure at these temperatures and current densities would occur within hours. It also establishes that these materials are more stable than metal features, paving the way toward their integration with operational CMOS.

## I. INTRODUCTION

Devices such as FinFETs and gate all around transistors have reached diminutive sizes where key length scales span a countable number of atoms, posing new challenges. For example, effects like tunneling can be sensitive to detailed atomic arrangements, and variations in the number and position of dopants can play an outsized role. In this context, a platform that enables integration of atomically precise devices with CMOS technologies can perform an important pathfinding function – to identify future opportunities and limitations.

Atomic precision advanced manufacturing (APAM) is a method to add electrically active dopants to silicon by leveraging the chemical contrast between a clean and reactive bare silicon surface, compared to a passive hydrogen terminated silicon surface. The APAM process, detailed in Fig. 1, is used to add dopants to the silicon surface at densities much greater than the solid solubility limit – a fundamentally different way to add dopants to silicon compared to implanting and high temperature activation of dopants. The chemically attached phosphorus dopants form a 2D sheet of dopants, also called a delta layer, and can be patterned (Fig. 1d) with a variety of methods and at different length scales, from macroscale patterning using photolithography [1] down to atomic precision by patterning the hydrogen termination with a scanning tunneling microscope (STM) [2]. Thus, APAM provides atomically precise placement of dopants both laterally and out-of-plane, enabling fabrication of ultrashallow junctions.

Initially pioneered to create atomic-scale quantum devices operating only at cryogenic temperatures [2-5], APAM has recently been developed as a manufacturing platform to explore the future of digital electronics in silicon by working toward the integration of delta layer material into CMOS devices [6, 7]. This is an appealing platform, as APAM can build dopant devices in silicon at the absolute limit of fabrication – the atomic scale – and can evaluate novel devices at length scales that are increasingly becoming relevant in modern CMOS FinFET geometries. More immediately, the high density of active dopants may be used to improve contact resistance in future CMOS technology nodes. Room temperature operation of an APAM device has recently been demonstrated, which is a necessary prerequisite to any practical application [8].

However, several technical hurdles need to be overcome before APAM can be seamlessly integrated into CMOS. The first is driving the APAM fabrication thermal budget down to be compatible with standard CMOS budgets. Historically, APAM devices are created at high temperatures (>1200°C) driven by the need to produce pristine silicon surfaces, but recent advances have driven the APAM fabrication temperature down toward <850°C which are more compatible with CMOS fabrication [9], and opens the door to integration of APAM into existing CMOS devices.

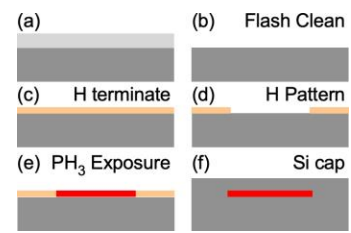

Fig. 1. The APAM process starting from passivated single crystal silicon in (a) that is cleaned via high temperature flash (~850°) to remove oxide and contaminants to produce a clean starting surface (b), then passivated with hydrogen in ultrahigh vacuum (UHV) (c), then patterned to expose bare Si (d), reacted with molecular species to form activated dopants (e) and finally capped by epitaxial Si in (f).

The second hurdle to APAM and CMOS integration is to understand the compatibility of the delta layer with CMOS architectures at relevant operational temperature and current conditions, which is the focus of this work. Back end of line compatibility has already been demonstrated by functional devices with a high-k dielectric/metal gate stack [10, 11]. In this paper, we report on our exploration of the compatibility of APAM with CMOS by conducting accelerated lifetime testing of APAM delta layer material at relevant current densities and show that delta layer material specifically is more robust than standard metal features in current CMOS devices, paving the way forward for APAM and CMOS integration.

II. EXPERIMENTAL DETAILS AND METHODS

A. Device Details

We chose an ASTM specified electromigration test structure [12] as our testing platform, as shown in Fig. 2(a). This four-terminal structure is typically used for metal-based electromigration testing, and thus is well suited to causing quick device failures from highly concentrated current densities. Four terminals are present in this test structure, two of which are used for applying a forcing current, and the other two for sensing the induced voltage drop. We used all standard dimensions from the ASTM guidelines except for the line widths We used linewidths of 5μm, 10μm and 20μm to serve as good analogs for more typical APAM structure geometries, with an 800μm long channel. These devices are placed adjacent to each other on a single silicon die.

Two varieties of the test structure layout were fabricated and tested. One version utilized a phosphorus ion implant in p-type silicon (referred to as P devices), and the second version utilized a phosphorus delta layer (referred to as P:δ devices). These two versions will allow a comparison between a conventional ion implant phosphorus layer to the ultra-dense, ultra-thin APAM material.

A series of both P and P:δ devices were prepared. The P devices' starting material was prepared by implanting a uniform phosphorus layer at 5 keV, 1e13 atoms/cm$^2$ dose with the implant centered around 0.03 μm deep into a wafer, which was then diced to produce implanted P device starting chips. The P:δ starting material was produced by an abbreviated APAM process to terminate the entire surface with phosphorus layer by skipping the hydrogen termination and patterning steps in Fig. 1(c-d). First individual 5 x 9 mm float zone silicon (100) chips were placed in a UHV environment, flash cleaning the surface at high temperatures (>850°C) to remove oxide and contaminants [Fig. 1(b)] and then exposing it to phosphine bled into the chamber at low pressures [Fig. 1(e)], which after a 15 min incorporation anneal at 300°C results in the creation of the highly doped phosphorus delta layer over the entire silicon surface. This P:δ 2D layer is then encapsulated by coating with epitaxial silicon with a process designed to minimize the diffusion of the P:δ. Silicon epitaxy was performed with a deposition rate of 0.5 nm/min. First, 2 nm of silicon were deposited via silicon arch evaporation without heating the substrate to form what is known as a locking layer to reduce subsequent phosphorus segregation [13]. The sample was then rapidly annealed in the vacuum chamber at up to 600°C for 15 seconds to recrystallize the locking layer. Next, the sample was brought to approximately 250°C for capping with an additional 30 nm of epitaxial silicon for devices or 50 nm for SIMS samples (to minimize surface effects) before removal from vacuum. This epitaxial cap is not perfect and has associated disorder and contamination but is compatible with device creation.

The delta layer has a dopant concentration of 2e14 cm$^{-2}$ [6, 14]. The implant and delta layer phosphorus distribution in the channel region of the device shown in Fig. 2(a) of equivalent samples was measured using secondary-ion mass spectrometry (SIMS) and shown in Figure 3. Since SIMS tends to broaden out the distribution due to forward scattering by the sputtering ions, we also indicate the "ideal" delta layer concentration of 5e20 cm$^3$ with an effective thickness of 4 nm, which is fitted to coincide with the peak P distribution in the SIMS data. The actual distribution of the delta layer will be between the ideal and SIMS measured distribution. The near surface data (<20nm deep), indicated by the vertical black dashed line, suffers from measurement artifacts and is not reliable.

Both types of chips then went through standard photo-lithography and ICP dry etch to define the mesa structure shown in dark gray regions of Fig. 2(a) with a cross section shown in Fig. 2(h). After etching, the aluminum bond pads were deposited which covered both the mesa structures and filling in etched hole contacts, shown in Fig. 2(b-e), to connect to the delta/implant layer with a large pad to allow wire bonding.

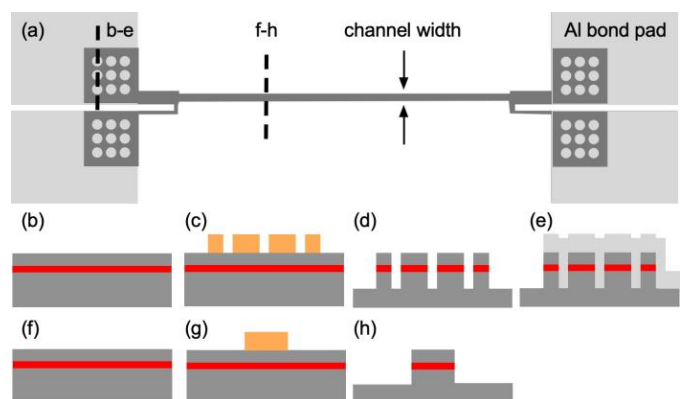

Fig. 2. (a) an overview of the robustness testing geometry with the dark grey regions indicating the mesa etch forming the conductive channel. Starting with a silicon chip with either a uniform delta layer or implant layer is then masked (c,g) and etched (d,h) to reveal the dark grey structure in (a). Contacts are made by depositing aluminum bond pads that contact the layers by filling contacts holes shown in cross section in (e). The vertical dashed lines refer to either the aluminum contact area created in (b)-(e) or the mesa etch created in (f)-(h).

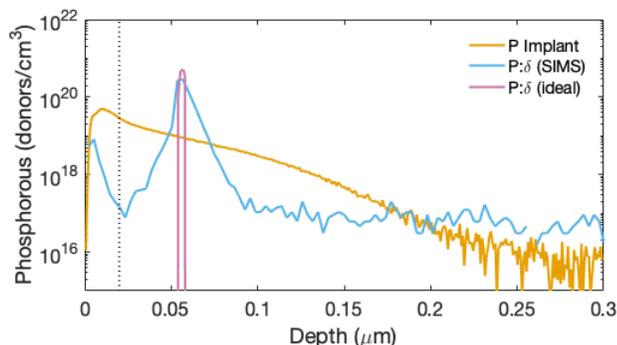

Fig. 3. Measured and ideal phosphorus concentration in the channel region of the P and P:δ layer lifetime device testers. The vertical black dashed line indicates the start of useful SIMS data.

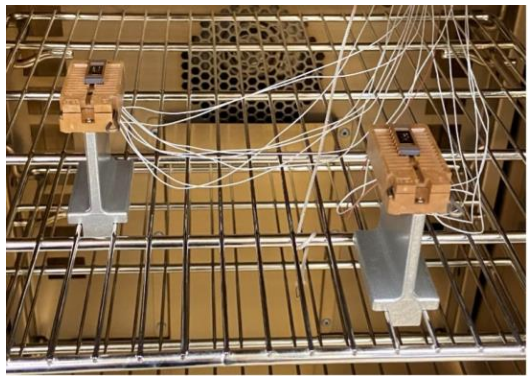

Fig. 4. Interior view of the testing chamber. ZIF sockets are elevated from oven grates using aluminum stands. Electrical connections between the ZIF socket and wire are made with crimp connections.

Each sample was placed into a high-temperature tolerant ceramic package and affixed with a high-temperature silver paste. Electrical contact was made between the device and package with 1-mil thick gold bond wires. The packages were left without a cover to facilitate easy removal after testing.

A zero-insertion-force (ZIF) socket was used inside the oven to allow quick device changes. Solder was completely avoided in all connections due to the high oven temperature. Instead, crimp and friction fit connectors were used where needed. A port at the rear of the oven allowed all connections to pass through. Fig. 4 shows the testing setup inside the oven.

### B. Device Test Conditions

The samples were brought to 300°C and allowed to equalize before biasing the drive terminal terminals. We drove each channel at a fixed current level of 30 mA, 60 mA, and 90 mA, for the 5μm, 10μm, and 20μm respectively. Driving the wider 20μm channel at high current caused failures in the bond wires within a few hours. Assuming a 200 nm thickness for the P and P:δ layers, the approximate minimum current density is 3.0 MA/cm$^2$ for the 5μm and 10μm channel width, and 2.25 MA/cm$^2$ for the 20μm width.

We utilized Keithley 2401 source-measure units in a constant-current mode to provide the electrical stress current for the test devices. Each unit was set up for remote control over a GPIB interface. Control and logging of the test setup was provided through a custom LabVIEW program which logged oven temperature, drive current, and the resulting voltage in the device under test at 30 second intervals for the duration of the test. The testers are designed to fail due to open-circuit or a percent-increase in resistance in the test line. In all tests, we ran the test features until the devices opened. When the devices begin to fail, the controllers attempt to drive the appropriate current by increasing the drive voltage up to a maximum of 20 V, at which point the device was considered open.

## III. RESULTS AND DISCUSSION

### A. Results Overview

We found that the narrower APAM P:δ and implanted P devices can survive the high current densities for several weeks at 300ºC as shown in Fig. 5. The time to failure for the implanted devices ranged from a few hours for the 20 μm device to over 70 days for the 5 μm device. At the end of this tests, all three implanted phosphorus samples failed, while only 1 of the delta layer devices failed by the end of the testing at day 76. The failures seen were an instantaneous change to an open circuit in all cases. Except for the 5 μm implanted channel, shown in red in Fig. 5(a), most samples showed steady resistance values until the devices failed. We attribute this variable behavior to likely a combination of electromigration of material in the bond wire and to Kirkendall voiding at the Au/Al interface [15] which caused variable resistance bouncing between a higher value and lower base line value until the device eventually failed.

After cooling to room-temperature, we resistance tested each structure to confirm the failures as seen by the software.

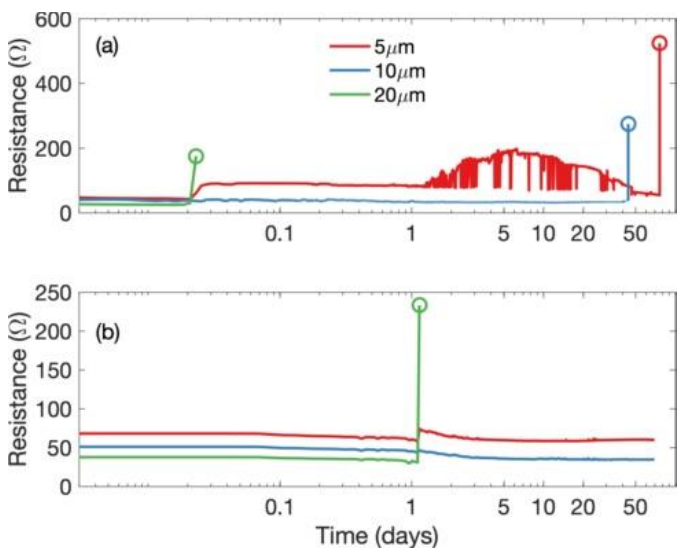

Fig. 5. Lifetime test data at 300ºC showing the resistance of the test structures as a function of time for (a) the P samples and (b) the P:δ samples. Circles indicate the test structure developing an open circuit.

*B. Failure Analysis*

After completion of accelerated lifetime tests, room temperature electrical probing and electron microscopy indicated that degradation of the metal bond pads and metal wires rather than the doped silicon was the cause of device failure. The physical die was separated from the ceramic packages and placed on a manual probe station to determine whether an electrical path was still present between contacts. We utilized four-terminal resistance measurements to look for continuity in the test structures. Measuring an as-fabricated untested P:δ sample as a control, the different robustness test structures have resistances of 30 kΩ for the 20 μm width up to 90 kΩ for the 5 μm width between the bond pads of the device. The measured resistances between bond pads of different robustness testers on the same silicon die (e.g. between the bond pads of the 5 μm and 20 μm testers) measure >125 kΩ indicating that the robustness structures are electrically connected through the die silicon as expected.

We measured electrical continuity in all the post-failure devices, both P:δ and implant P samples, indicating that the failures were not in the line test structures themselves, but rather the bond wires. Across all the individual resistances of the test structures tested (3 implanted P, 3 P:δ delta layers) the post test resistance values varied between 45kΩ and 220 kΩ in seemingly arbitrary ways. These variations are likely due to the competing influence of electromigration and drift of metal species into the region below the bond pads, which improved the contacts, and simultaneous degradation of the contact pads due to oxidation and disappearing aluminum, which increased the resistance values. Similarly, contact between bond pads of different robustness testers also varied with most resistance values >100 kΩ.

We examined the failed devices using scanning electron microscopy (SEM) in a Helios G4 PFIB from Thermo Fisher Scientific. All SEM images were collected using 5 kV accelerating voltages and imaging current of 0.8 nA. Shown in Fig. 6(a) is a representative region of the bond pad with the bond wire still connected. We observed the electromigration of the gold bond wires shown in the "bamboo" wire structure shown in the section of the wire on the left of Fig. 6(a). We can also see significant gold transfer to the bond pads, shown in the bright dendritic areas around the bottom of the bond pads, confirmed as gold with energy-dispersive X-ray spectroscopy (EDS) (not shown). This indicates that the failures measured during the accelerated testing is exclusively in the bond wires, and not in that delta layers or implanted sections considering both electrical continuity data and observed bond wire damage.

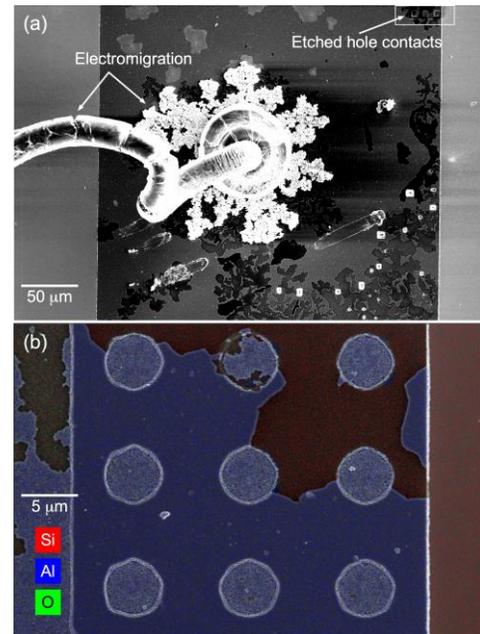

Fig. 6. (a) SEM image of delta layer bond pad. The necking down pattern in the bond wire indicates electromigration. Oxidation and gold transfer to the bond pad are also apparent. The box in upper right highlights the etched hole contact area. (b) A close up EDS map of the etched hole contact region in a different area than (a) showing the disappearance of the aluminum.

The bond pads showed regions of missing aluminum likely indicating beginning the intermixing of the aluminum bond pad and underlying silicon which occurs at temperatures >200°C [16]. This is shown in Fig. 6(b), where regions of the aluminum have disappeared. There was no evidence of any phosphorus accumulation on the surface.

*C. Modeling Results*

After determining that failure was confined to the bond wires and that the line testers were still conductive, we modeled the high temperature current paths in both device types to confirm that the current at elevated temperatures was flowing through the P and P:δ instead of through the underlying silicon.

A simplified cross sectional 2D model was developed for both device types to determine where current was flowing in the vertical direction is shown in Fig. 7(a), as well as the finite element current map near the contact. Sandia's own open-source semiconductor device modeling code suite, Charon, was used in these simulations [17, 18]. Within the simulation, we used a constant bias across the model up to 0.5 V at various temperatures up to 300°C to determine the current paths through the sample. We simulated the phosphorus concentration as taken from the SIMS data as shown in Fig. 3, but ignoring the unreliable surface data <20nm indicated by the dashed black line. The handle silicon doping was 1e16 cm$^{-3}$ acceptor doping [11], while the doping of the epitaxial cap in the P:δ samples was 5e18 cm$^{-3}$. The device with an implanted P layer was simulated using a triangular mesh with the smallest mesh size of 7nm, since the doping profile (Fig. 3) is smooth allowing for relatively big mesh size. The devices with a P:δ layer (using either the SIMS doping profile or the ideal profile) were simulated using rectangular mesh with a very small mesh size of 0.01 nm at the delta layer along the direction perpendicular to the delta layer due to high doping and strong electric field there. The temperature dependence of the silicon mobility was handled by the Arora material model [19].

It's noted that the channel in the actual devices is 800 μm long. In the simulation, in order to adequately resolve the highly doped delta layer, very small mesh size was needed, so it was impractical to simulate the entire 800 μm length. Instead, we chose to simulate a 2-μm long device between the contacts and use the simulated current to determine a resistance for the 2-μm channel. We then scaled the simulated resistance by 400 to compare with the full 800-μm channel in the experiment. The measured experimental currents on these devices are on the order of 10 mA at 0.5 V bias. Without any calibration work, our scaled simulated current ranges from about 5 mA for the P implant to 30 mA for the P:δ SIMs at 0.5 V, which is on the same order as the measured value, giving us confidence our model matches physical reality.

Figure 7(b) shows the current density as a function of sample depth along the vertical line indicated in Fig. 7(a). The simulations show that in both cases the current is tightly confined near the surface. For the three cases, 95% of the current is confined within 142 nm of the surface for the implant, within 58 nm for the ideal P:δ, and 88 nm for of the SIMS P:δ. Using these confinement values, the current density in the implant P layer is 3.2 MA/cm$^2$ for the 5μm and 10μm and 4.2 MA/cm$^2$ for the 20 μm channel. For the P:δ layer, the current density is estimated between 5.1 and 7.8 MA/cm$^2$ for the 20μm channel, and between 6.8 and 10.3 MA/cm$^2$ for the 5 μm and 10 μm channel using the ideal and SIMS confinement values. These values exceed the current capacity of copper (~1 MA/cm$^2$) at the same temperature (300°C) [20], and possibly competitive with multilayer CVD-grown graphene interconnects with maximum room temperature breakdown values of ~40 MA/cm$^2$ [21].

This simulation work confirms that at elevated temperatures the current is confined to the phosphorus layers in all samples, indicating that our experimental conclusions that the APAM P:δ is more robust than the metal features in CMOS and there are no barriers to future CMOS integration are valid.

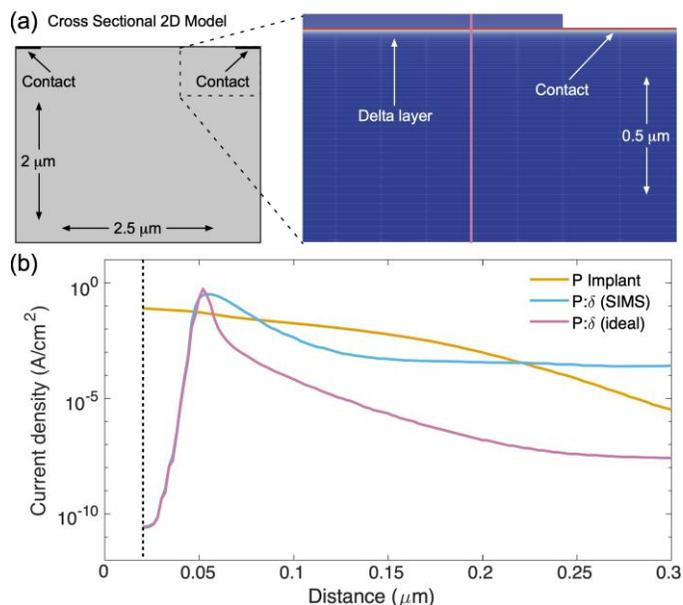

Fig. 7. (a) The cross sectional 2D model showing the finite element mesh and (b) the simulated current flow in a vertical slice through the channel area shown by the vertical line in (a) for the P layer in the implanted, SIMS, and ideal P:δ layers showing the current density as a function of distance into the silicon. Current is largely confined to the layer regions and is negligible through the deep substrate. Unreliable SIMS data <20nm deep indicated by the dashed black line was ignored.

IV. CONCLUSION

In this work, we use accelerated lifetime tests of P:δ delta-layer APAM devices at high temperature and current and establish that these delta-layer materials are robust and compatible with CMOS and that there are no fundamental barriers related to lifetime compatibility to having APAM elements integrated into operational CMOS. We showed that the failures were not the phosphorus layers, but instead in the metal wire bond interconnects used to package the device. We verified through simulation that the current at 300°C is flowing exclusively through both the P:δ and P implant layer, validating the experimental results. Additionally, the current densities and stressing temperatures used are well in excess of what would be seen under normal operation in CMOS. These experiments show that the APAM delta layer structures are more robust than the wire bonds used to make device contacts, clearing any lifetime issues as a barrier to the practicality of combining APAM into CMOS devices. Current ongoing work is attempting to add APAM to CMOS in a temperature compatible way to preserve both CMOS and APAM function in the final integrated device.